\documentclass[preprint2,twocolumn]{aastex63}
\usepackage{color}
\usepackage{url} 
\usepackage{hyperref} 
\usepackage{graphicx} 
\usepackage[latin1]{inputenc}
\usepackage[T1]{fontenc}
\usepackage{amsmath}
\usepackage{amsfonts}
\usepackage{amssymb}
\usepackage{booktabs}
\usepackage{array}
\usepackage{float}
\usepackage{natbib}
\usepackage{booktabs}
\usepackage{amsmath}
\usepackage{array}
\usepackage{wasysym}
\newcommand{\ecs}{erg~cm\pwr{-2}~s\pwr{-1}}

\newcommand{\ebv}{$E(B-V)$}

\newcommand{\pwr}[1]{$^{#1}$}

\makeatletter
\newcommand*{\rom}[1]{\expandafter\@slowromancap\romannumeral #1@}
\makeatother
\makeatletter
\def\blfootnote{\xdef\@thefnmark{}\@footnotetext}
\makeatother

\newcolumntype{R}[1]{>{\RaggedLeft\arraybackslash}p{#1}}
\newcolumntype{C}[1]{>{\centering\arraybackslash}p{#1}}

\received{--}
\revised{--}
\accepted{--}
\submitjournal{PASP}
\shorttitle{An AGB Star with a Thick Circumstellar Shell}
\shortauthors{Mould et al.}

\begin{document}
\correspondingauthor{Jeremy Mould}
\email{jmould@swin.edu.au}
\title{An AGB Star with a Thick Circumstellar Shell}
\author[0000-0003-3820-1740]{Jeremy Mould}
\author[0000-0002-2126-3905]{Mark Durr\'{e}}
\affil{Centre for Astrophysics \& Supercomputing, Swinburne University, PO Box 218, Hawthorn, Vic 3122, Australia}
\author[0000-0002-9413-4186]{Syed Uddin}
\affil{Carnegie Observatories, 813 Santa Barbara St, Pasadena, CA, USA}
\author{Lifan Wang}
\affil{Mitchell Institute for Fundamental Physics and Astronomy, Texas A\&M University, College Station, TX 77843, USA}
\affil{Purple Mountain Observatory, Nanjing, 201008, Jiangsu, People's Republic of China}
% UNCOMMENT THE LINES BELOW IF YOU WISH TO USE BIBTEX
%Citations may be made using the natbib commands \citet{},\citep{} etc.
%\usepackage[authoryear]{natbib}
%\bibpunct{(}{)}{;}{a}{}{,}
%\setlength{\bibsep}{0.3mm}
%\usepackage{aas_macros}

%\hypersetup{colorlinks,citecolor=blue,linkcolor=blue,urlcolor=blue}
\newcommand{\WJ}{WISEA J173046.10\nobreakdash-344455.5}
\newcommand{\WJS}{J173046}
\newcommand{\TBB}{1305~K}
\begin{abstract}
The Asymptotic Giant Branch (AGB) is the terminal phase of red giant evolution with timescales of millions of years and a total mass lost from the star that is a significant fraction of the initial mass. Investigation of one of these stars, \WJ, a kpc in the direction of the center of the Galaxy, reveals a cool oxygen rich star with a dust shell of black-body temperature \TBB.
\end{abstract}

\section{Introduction}
It will soon be 50 years since astrophysicists first evolved stellar models of low and intermediate mass stars beyond core helium burning to the double shell source phase. They encountered a number of phenomena not seen in simpler structures: thermal pulses and convective troughs which mixed burning products to the surface, terminal mass loss leading to the exposure of hot cores (and planetary nebulae), a powerful observational constraint in the form of the initial final mass relation, an instability region in the cool HR diagram and a rapid luminosity evolution (1 mag per million years).

Unsurprisingly, infrared observations have proved illuminating of this phase of evolution. The Magellanic Clouds have proved a valuable laboratory \citep{Meixner2008,Cioni2004,Frogel1990}. Much complexity remains to be explored, and \textit{Gaia}'s Galactic distances can be expected to provide a solar metallicity contrast with what has been discovered there. 

Among the problems still challenging AGB evolution are hydrodynamic treatment of stellar and circumstellar structure, single star loss winds and binary mass transfer. AGB stars are the stellar site for the main s-process (slow neutron capture process), with $^{13}$C being the major neutron source via $^{13}$C($\alpha$,n)$^{16}$O \citep{Gallino1998,Karakas2019}, with challenges to develop a quantitative treatment. We can expect progress with the advance of more powerful observations at high spatial and spectral resolution.

\begin{figure}
	\centering
	\includegraphics[width=1\linewidth]{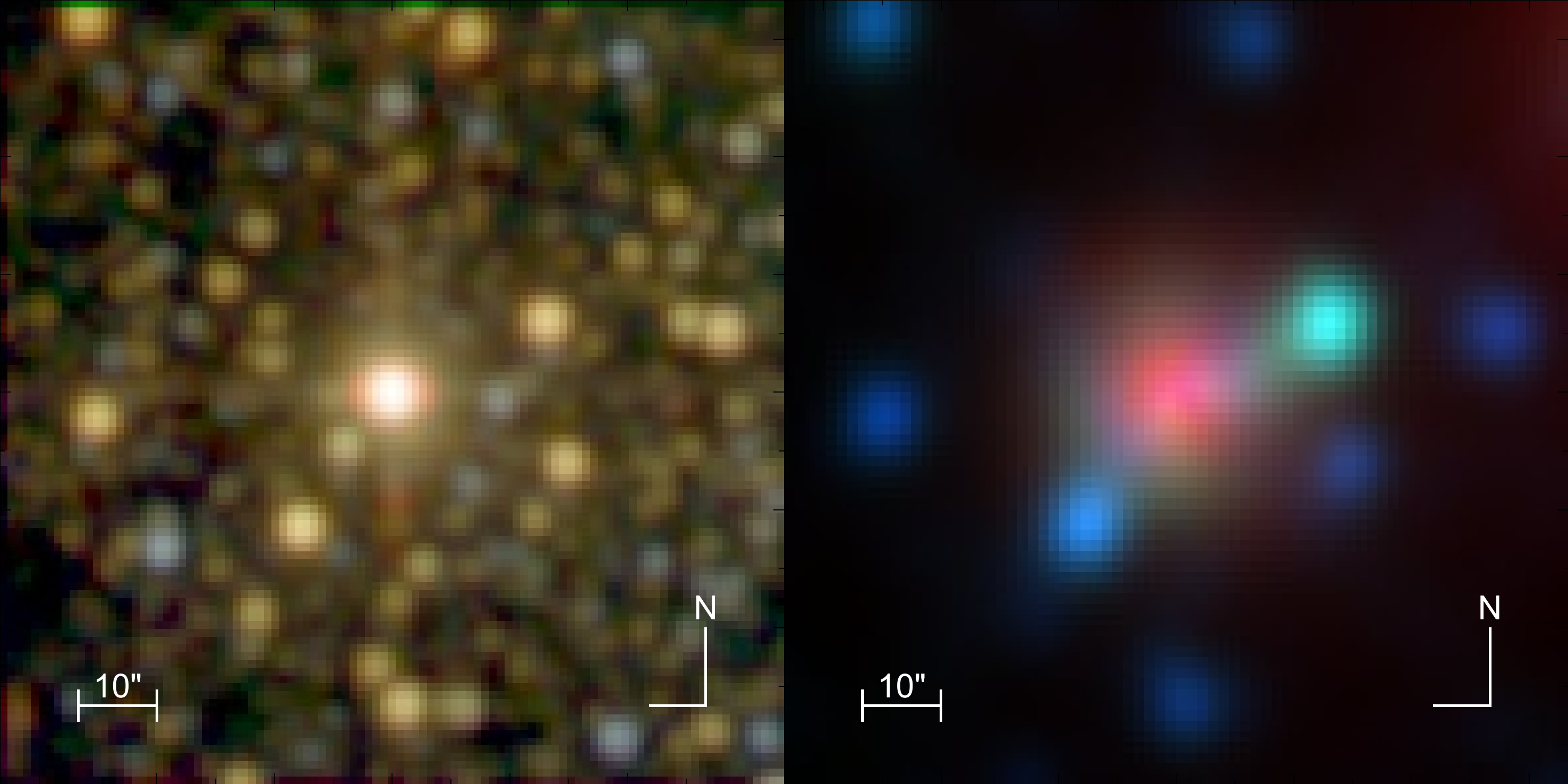}
	\caption{Color images of \WJ. Left: 2MASS $J/H/K_{s}$ (spatial resolution 1\arcsec/pixel - logarithmic scaling). Right: WISE $W1/W2/W3$ (spatial resolution 1.375\arcsec/pixel - linear scaling). Colors for each filter are blue/green/red correspondingly. The length scales and orientations are shown. The object is significantly redder than field stars.}
	\label{fig:at2019gac07}
\end{figure}

The optical transient AT2019gac was discovered on 27 May, 2019 by the MASTER Global Robotic Net (MASTER OT J173045.95-344454.7)\footnote{\url{https://wis-tns.weizmann.ac.il/object/2019gac}}, with a discovery magnitude of 16.5. Rather than discard it as a non-supernova on the basis of the spectrum, we pursued it as a long period variable (LPV) star. This was spatially identified with \WJ{} (2MASS J17304612-3444551), henceforth \WJS. The 2MASS and WISE color images are shown in Figure \ref{fig:at2019gac07}, illustrating the redness of the object.
\
Infrared spectroscopy and photometry show that it is a luminous asymptotic giant branch (AGB) star with a large, thick, and cool circumstellar shell, located in a region a kpc from the Sun towards the Galactic Centre, that has been actively forming stars in the last 0.1 to 1 Gyr.

\section{Optical Spectrum and Distance}
\begin{figure}[!htpb]
\includegraphics[width=1\linewidth]{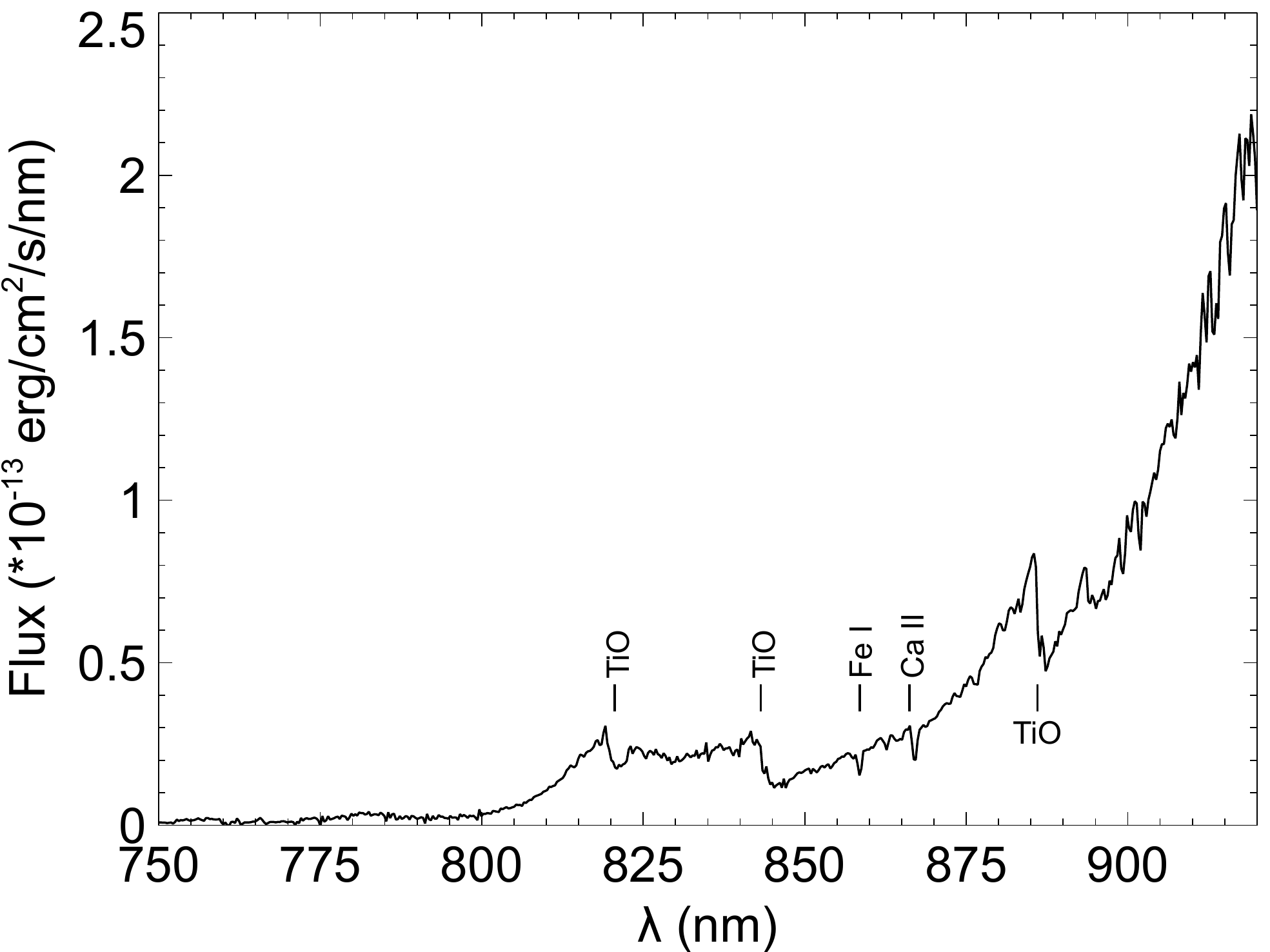}
\caption{WiFeS spectrum of AT2019gac. The spectrum is a late M star, showing TiO bandheads and atomic absorption lines.}
	\label{fig:at2019gacws}
\end{figure}

\WJS{} has a very red color \citep[$i-z$ = 3.84 from SkyMapper data;][]{Wolf2018}. SkyMapper $z$ images also show a decline in brightness of $\sim 0.5$ mag over the period 1 August, 2014 to 14 March, 2018. The optical spectrum was obtained on June 13, 2019 with the WiFeS spectrograph \citep{Dopita2007} on the ANU 2.3m telescope at Siding Spring Observatory. A standard star was observed (Feige 110), the data were reduced using PyWiFeS software \citep{Childress2013}, and the spectrum (longward of 750 nm) is shown in Figure \ref{fig:at2019gacws}. Shortward of this, the spectrum is non-existent at the flux limit.

%Synoptic photometry was obtained by ASAS-SN \citep{Jayasinghe2019} and is shown in Figure \ref{fig:at2019gaclc}.
%
%\begin{figure}[h]
%\includegraphics[width=1\linewidth]{AT2019gac_04}
%\caption{$g$ band ASAS-SN photometry (7 March -- 15 June, 2019). The light curve shows magnitude level variability with a hint of periodicity. The arrows mark observation dates as follows: 1 - transient observed (27 May 2019) 2 - optical spectrum (13 June 2019) 3 - infrared spectrum (25 August 2019)}
%	\label{fig:at2019gaclc}
%\end{figure}
%
\WJS{} has also been observed by ESA's \textit{Gaia} satellite \citep{TheGaiaCollaboration2016}. Its parallax in DR2 is $0.921 \pm 0.258~{\rm mas}~=~1086(+422,-238)$ pc. \citep{TheGaiaCollaboration2018,Bailer-Jones2018} We neglect the DR2 parallax offset of --0.08 mas \citep{Stassun2018}. The source catalog gives passband magnitudes of $m_G~=~16.07 \pm 1.1$ and $m_{RP}~=~14.1 \pm 1.05$, with the variability derived from the mean error in the fluxes. \WJS{} is not flagged as a variable in Gaia DR2; however, the are 135 g band magnitude measurements yielding a 0.066 mag standard error on the mean. If it is a LPV, more observations are required to find a period.

\WJS{} is located in a region with a number of young star clusters: Collinder 133, ESO 392-13 and Ruprecht 126 \citep{Kharchenko2013}. Their distances from the Sun range from 0.8 to 2.3 kpc, their reddening from $E(B-V) = 0.11-0.82$ mag, and their ages from 0.9--8 $\times$ 10$^8$ years.

\section{IR Spectroscopy \& Photometry}
\subsection{Spectroscopy}
\begin{figure}[h]
	\centering
	\includegraphics[width=1\linewidth]{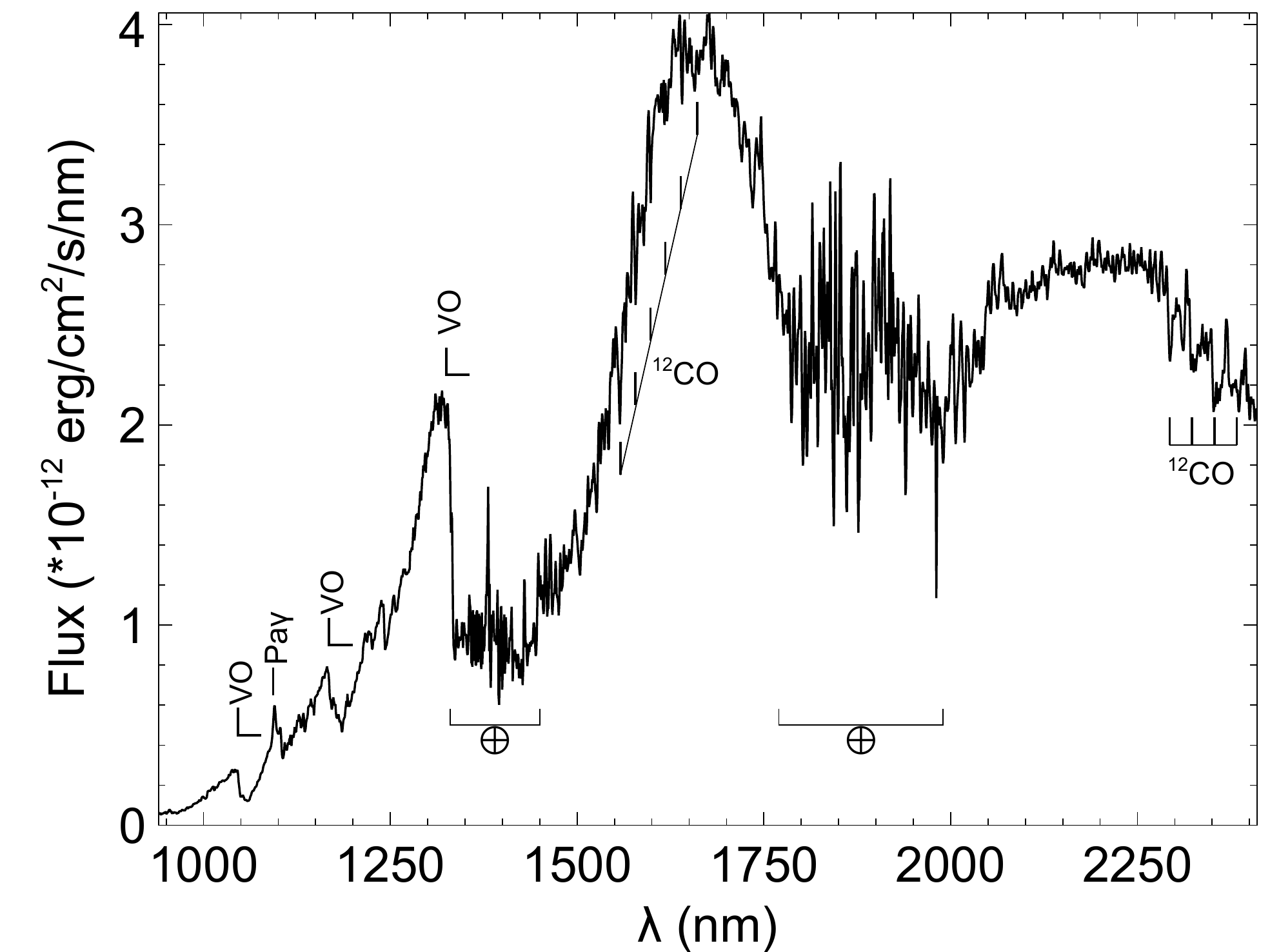}
	\caption{NIR Spectrum of AT2019gac. Note the noisy regions between the atmospheric windows, designated with $\oplus$ symbol (1330-1450 nm and 1770-1990 nm). Spectral features include the VO bandheads in the \textit{J} band, plus CO in \textit{H} and \textit{K} bands. Pa$\gamma$ is identified in emission.}
	\label{fig:at2019gac01}
\end{figure}

The steep rise in Figure \ref{fig:at2019gacws} demands an infrared spectrum. This was observed  with the SofI (Son of ISAAC) near infrared (NIR) spectrometer \citep{Moorwood1998} at the Nasmyth A focus of the ESO 3.5-m New Technology Telescope (NTT) at the European Southern Observatory (ESO) in La Silla (Chile), as a commensal observation in the program 0103.B-0504(B) (observation date 25 August 2019). Both the GBF and GRF grisms were used with a 1\arcsec{} slit width, for a spectral range of 918\nobreakdash-2518 nm and a spectral resolution $R\sim600$. The A0 star HIP79473 was also observed as a telluric standard and flux calibrator. The data were reduced in the standard manner, with flat-fielding, extraction of the spectrum from the A-B pair observations, wavelength calibration, and combination of all frames for each object. Telluric correction and flux calibration were performed using an A0V stellar spectral template \citep[from the ESO stellar library of][]{Pickles1998}. The GBF and GRF spectra were combined, showing excellent flux calibration at the boundary wavelength (1600 nm), as seen in Figure \ref{fig:at2019gac01}. The spectrum was re-dispersed to the range 940-2410 nm, with 0.5 nm steps.

\subsection{Photometry}
\begin{figure}[h]
	\includegraphics[width=1\linewidth]{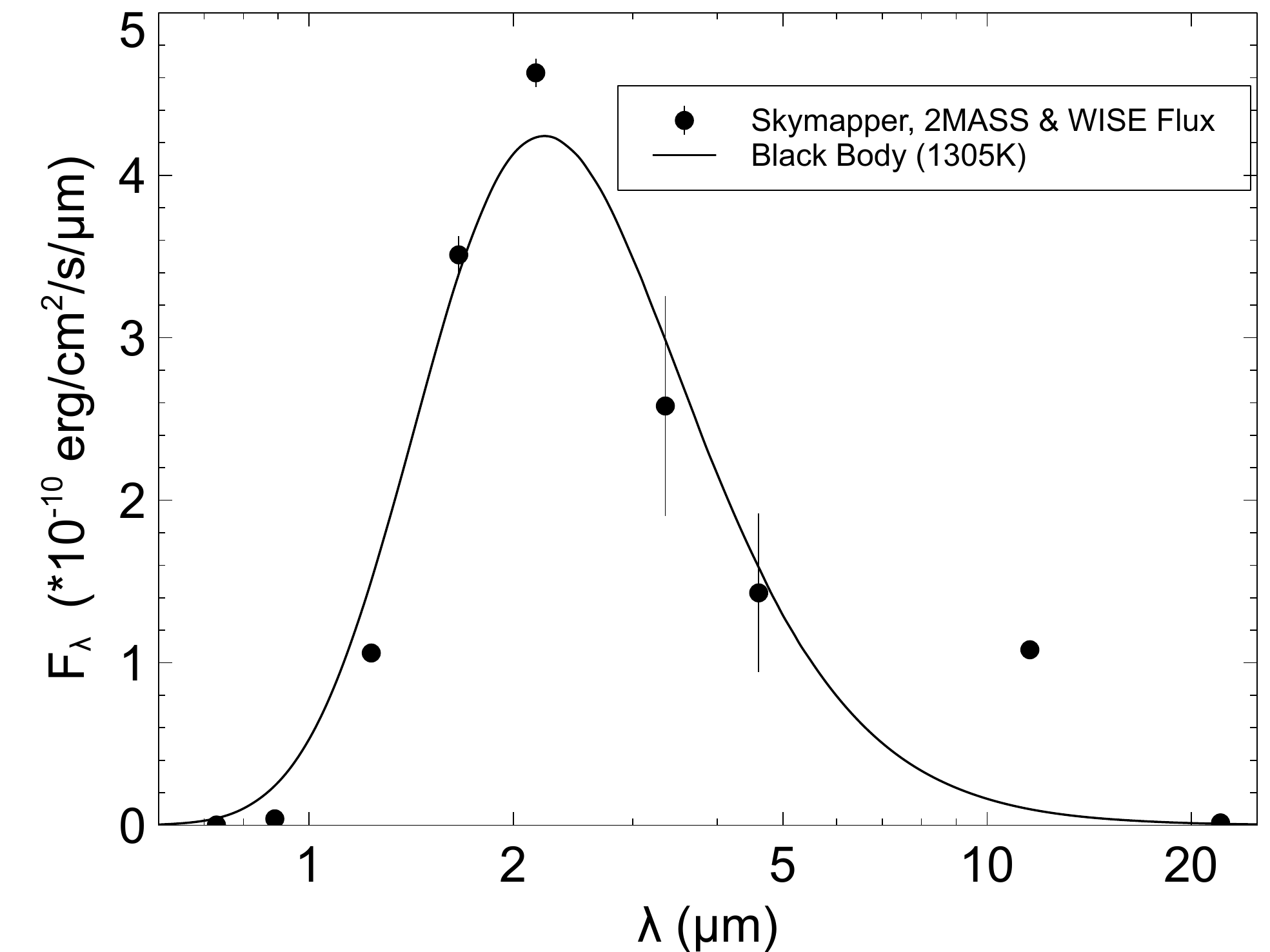}
	\caption{Spectral energy distribution from SkyMapper, 2MASS and \textit{WISE} photometry. The flux units are 10$^{-10}$ erg cm$^{-2}$ s$^{-1}$ $\mu$m$^{-1}$. The curve is a \TBB~ black-body.}
	\label{fig:at2019gac_sed}
\end{figure}

\WJS{} has also been observed by NASA's \textit{WISE} satellite \citep{Wright2010,Cutri2013}, and has 2MASS photometry \citep{Skrutskie2006}. Combining these with SkyMapper \textit{i} and \textit{z} data, we can plot the spectral energy distribution (SED), shown in Figure \ref{fig:at2019gac_sed}. A black-body curve was fitted to the photometric points, yielding a best-fit temperature of \TBB. The excess flux at the WISE $W3$ band can be attributed to silicate emissions at $\sim10\mu$m.

Computing the integrated flux of the fitted black-body, we obtain $F_{BB}=1.47\times10^{-8}$ \ecs. Equating the luminosity of a black-body of radius $R$ and temperature $T$ with that derived from the observed flux at distance $D$ (found above), we obtain the relationship:
\begin{equation}\label{eqn:eq1}
R~=~\frac{D}{T^{2}}~\sqrt{\frac{F_{BB}}{\sigma}}
\end{equation}
where $\sigma$ is the Stefan-Boltzmann constant. 

From this we derive the radius of the dust shell $R=455 (+177,-101)~\nom{R}$ and luminosity $L=540 (+1340,-210)~\nom{L}$ (the uncertainties are from the range of \textit{Gaia} distances). Given the solar bolometric magnitude of 4.75, this equates to $M_{Bol}=-2.1 (+0.6,-1.3)$.

\section{Discussion}
\subsection{Spectral Type and the Dust Shell}
The features of the spectrum (the broad bumps in the $H$ and $K$ spectral regions and the molecular features in the $J$ band) are similar to a late-class M star, however with a different general slope. Inspecting the IRTF IR spectral library of cool stars \citep{Rayner2009}, we noticed a striking similarity to IRAS01037+1219 (WX Psc). This is classed as a variable of type M~ (OH/IR), i.e. a Mira at the very late stage of AGB evolution, with a highly dusty shell, extreme mass loss rate and OH maser emission. A search of the 1612 MHz OH blind survey of the galactic bulge region by \cite{Sevenster1997}\footnote{\url{http://cdsarc.u-strasbg.fr/viz-bin/cat/J/A+AS/122/79}} did not show a match.

Full treatment of the emission and absorption of the star and shell would require radiative transfer code like DUSTY \citep{Ivezic1999}, but this is beyond the scope of this paper. To determine the extinction (and thus the stellar magnitude), we can model the spectrum using a late M stellar template plus a standard dust extinction curve. We also include a contribution from the black-body emission modeled by the SkyMapper, 2MASS and WISE data. Our model is thus:
\begin{equation}\label{eqn:eq2}
F_\lambda~=~\alpha~T_\lambda~a_\lambda + \beta~ B_\lambda 
\end{equation}
where $F_\lambda$ is the observed spectrum, $T_\lambda$ is the template spectrum, $B_\lambda$ is the black-body contribution and $a_\lambda$ is the wavelength-dependent absorption from the circumstellar shell. $\alpha$ and $\beta$ are arbitrary scaling constants to be fitted. 

We selected 8 stellar templates from the IRTF library with spectral type M7III to M9III. We modeled the extinction using both the \cite{Cardelli1989} (CCM) and the \cite{Calzetti2000} (CAL) curves, with  the black-body temperature fixed at \TBB, as derived from the SED fit. The best fit was with the template IRAS21284-0747 (HY Aqr), spectral type M8-9III. Both forms of the extinction curve gave similar results; the CCM extinction model gave sightly better fit, with an \ebv~=~3.06. This translates to a \textit{V} band extinction $A_V \approx 9.5$ mag (using the standard CCM factor $R_V = 3.1$). Allowing the black-body temperature to vary in the fit produced almost no change. 

\begin{figure}[!htbp]
	\centering
	\includegraphics[width=1\linewidth]{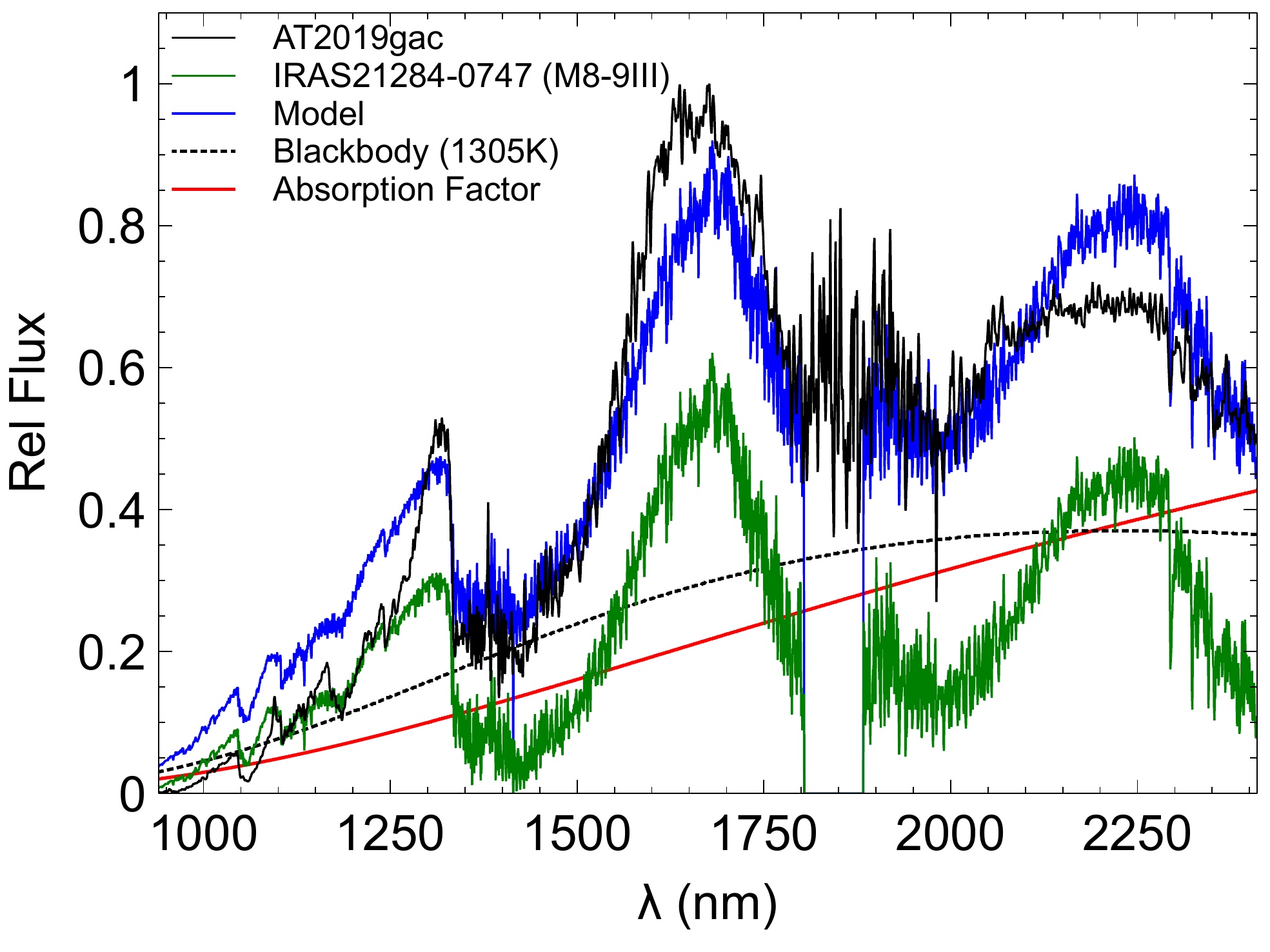}
	\caption{Model of a M8-9III stellar template fit to the IR spectrum of \WJS{} with an extinction \ebv=3.06 (CCM law), plus a contribution from the circumstellar shell black-body emission at \TBB. The plot shows the observed spectrum (black), the template spectrum (green), the circumstellar shell absorption factor (red) and the black-body contribution (dashed black), with the best fit model (blue). Fluxes are in arbitrary units and normalized.}
	\label{fig:at2019gac02}
\end{figure}

The plot of this fit is shown in Figure \ref{fig:at2019gac02}. The fit broadly reproduces the features of the NIR spectrum; the broad peaks in the \textit{H} and \textit{K} and the molecular features in the \textit{J} band combined with the steep decline in flux towards shorter wavelengths. The detailed flux levels are somewhat different, however this can be ascribed to (a) the differences in extinction between the dust shell and the CCM model and (b) differences in detail between the template and the stellar spectra. The circumstellar shell and the stellar emission contribute roughly equal amounts of flux over the NIR range.

At the effective wavelength of the 2MASS $K_s$ filter (2159 nm), we fitted an extinction factor of 0.37 ($=1.1$ mag); with the range of \textit{Gaia} distances the 2MASS magnitude ($K_s~=~4.92\pm0.02$) translates to $M_K~=~-6.15\pm0.6$.
%\begin{table}[!htbp]
%	\centering
%	\caption{Spectrum Model Results}
%	\label{tbl:AT2019gac_Model}
%	\footnotesize
%\begin{tabular}{llrc}
%	\toprule 
%	Extinction& Stellar Template& Extinction & black-body \\ 
%	Law& (Type)&\ebv&Contribution\\
%	\midrule
%	CCM & BRI B1219-1336 (VX Crv)& 3.46 & No \\ 
%	&(M9III)&&\\
%	CAL & HD10884 (BK Vir)& 2.03 & Yes \\ 
%	&(M7III)&&\\
%	\bottomrule 
%\end{tabular} 
%\end{table}
\subsection{Comparison with Surveys}
The most comprehensive study of cool star SEDs is the SAGE survey of the Magellanic Clouds \citep{Meixner2006}. \cite{Woods2011} assigns 13 classifications to point sources, and \WJS{} fits O-AGB better than a Young Stellar Object (YSO) in this system; that is to say, it is an oxygen rich (M type) AGB star at the faint end of Woods's bolometric luminosity function.
Without a firm period measurement, it is not possible to tell to which of four sequences of  \cite{Wood2000} of variable red giants \WJS{} belongs.

We can also compare this object to the study by \cite{Wood2015} of variability of luminous red giants in the Large Magellanic Cloud (LMC). The Weisenheit reddening free magnitude of the object is 
\begin{equation}
W_{JK}=K~-~0.686~(J-K)
\end{equation}
This has a value of 2.33; at the distance of the LMC ($m-M=18.45$), this translates to $W_{JK}=10.8$, which is within the range for Wood's Figure 10 plot. Since there is an unknown contribution from the LMC's stars' circumstellar shells, we can only say that $W_{JK}$ and $K$ are plausible for AGB LPVs.
\subsection{Mass Loss}
Although we tend to associate thick dust shells with the terminal maximum luminosity phase of AGB evolution, it is quite possible from the point of view solely of the mass budget for a low luminosity AGB star to supply the requisite dust. A low or intermediate mass star may lose as much as 0.3 \nom{M}~ of gas at the time of the helium flash; after that the  \cite{Reimers1977} mass loss rate for a one solar mass star with the parameters we find here is of order $10^{-5}$ \nom{M} yr\pwr{-1}. \cite{Lebzelter2005} have studied variable red giants in 47 Tuc with a luminosity $\log L/\nom{L}=3.15$, considerably below the tip of the red-giant branch (RGB) at $\log L/\nom{L}=3.35$. Models that have undergone mass loss reproduce observed period-luminosity relations and they show that mass loss of the order of 0.3 \nom{M} occurs along the RGB and AGB. They show that stars evolve up the RGB and first part of the AGB pulsating in low order overtone modes, then switch to fundamental mode at high luminosities.

Assuming an dust absorption coefficient of 0.4 cm$^2$ gm$^{-1}$ \citep{Chini1991} and a gas to dust ratio of 100, we can compute the gas mass required to give $A_V=9.5$ mag; this is $M_{Gas}=4.9\times10^{-5}\nom{M}$. Thus there is ample dust to provide an optically thick shell.
\subsection{Possible Origin of the Transient}
The transient was at least 1.5 mag visually brighter than its minimum (given the limiting magnitude of 17.9 for the MASTER network), but had faded over the 90 day period between discovery and our IR observation (our estimate from the IR spectrum is $K_S=5.9$ mag, below that of the 2MASS catalog of 4.92 mag). The cause of the transient is not known; this could be intrinsic, e.g. the prototypical AGB LPV (Mira) has been shown to have X-ray flares on a 1-2 day timescale \citep{Karovska2005} and the symbiotic Mira V407 Cyg produced a nova outburst \citep{Munari2011}. Extrinsically, the dust shell may not be uniform \citep[e.g.][]{Tatebe2008,Paladini2012}, possibly caused by companions, stellar asymmetry, asymmetric dust emission or dust clumps. 
 
\section{Conclusion}
The transient AT2019gac is identified with the very red object \WJ. The color is due to the late spectral type of the central red giant with some foreground reddening, but notably to a circumstellar shell of warm dust. This shell has an estimated temperature of \TBB, a luminosity of 540 \nom{L} and a radius of 455 \nom{R}. From the observed extinction, the gass mass is $4.9 \times 10^{-5} \nom{M}$. On the basis of our observations, \WJS{} is a mass losing, LPV, AGB star, whose initial mass would have been comparable to the turnoff masses of the star clusters of its region in the Galaxy.

With a WISE W4 mag of 1.34, \WJS{} would repay spectroscopy at 3--30$\mu$m to investigate this shell in detail. We are only now beginning to get a good sample of Galactic LPV luminosities thanks to Gaia \citep[e.g.][]{Lebzelter2019}.

\acknowledgements

We thank the reviewer for their constructive report, which much improved the paper. This paper is based on observations collected at the European Organisation for Astronomical Research in the Southern Hemisphere under ESO program 0103.B-0504(B). We thank Purple Mountain Observatory, Nanjing, PRC for its support of this program.

This work presents results from the European Space Agency (ESA) space mission \textit{Gaia}. \textit{Gaia} data are being processed by the Gaia Data Processing and Analysis Consortium (DPAC). Funding for the DPAC is provided by national institutions, in particular the institutions participating in the Gaia MultiLateral Agreement (MLA). 

This publication makes use of data products from the Wide-field Infrared Survey Explorer, which is a joint project of the University of California, Los Angeles, and the Jet Propulsion Laboratory/California Institute of Technology, funded by the National Aeronautics and Space Administration (NASA) and from the Two Micron All Sky Survey, which is a joint project of the University of Massachusetts and the Infrared Processing and Analysis Center/California Institute of Technology, funded by the NASA and the National Science Foundation. 

The national facility capability for SkyMapper has been funded through ARC LIEF grant LE130100104 from the Australian Research Council, awarded to the University of Sydney, the Australian National University, Swinburne University of Technology, the University of Queensland, the University of Western Australia, the University of Melbourne, Curtin University of Technology, Monash University and the Australian Astronomical Observatory. SkyMapper is owned and operated by The Australian National University's Research School of Astronomy and Astrophysics. 

\facilities{ATT:WiFeS, NTT:SofI, Gaia, NEOWISE, SkyMapper}
\bibliography{library}	
\end{document}